\documentclass[aps,prl,twocolumn,superscriptaddress,showpacs]{revtex4}
\usepackage{latexsym}
\usepackage{graphicx}
\usepackage{epsfig}
\newcommand \be {\begin{equation}}
\newcommand \bea {\begin{eqnarray}}
\newcommand \ee {\end{equation}}
\newcommand \eea {\end{eqnarray}}
\newcommand \bed {\begin{displaymath}}
\newcommand \eed {\end{displaymath}}

\newcommand{\bit}{\begin{itemize}}
\newcommand{\eit}{\end{itemize}}

\begin{document}

\title{Non-commutative dynamics  and roton-like spectra  in bosonic and fermionic condensates}
\author{Paolo Castorina}
\email[]{paolo.castorina@ct.infn.it}
\affiliation{Dept. of Physics, University of Catania $\;$ and  $\;$  INFN, Sezione di Catania,\\ 
Via S. Sofia 64, I-95123, Catania, Italy
}
\author{Giuseppe Riccobene}
\email[]{riccobz@yahoo.it}
\affiliation{
Scuola Superiore di Catania, via S.Paolo 73, I-95123, Catania, Italy
}
\author{Dario Zappal\`a}
\email[]{dario.zappala@ct.infn.it}
\affiliation{INFN, Sezione di Catania,$\;$  and  $\;$ Dept. of Physics, University of Catania,\\
via S. Sofia 64, I-95123, Catania, Italy
}
\date{\today}
\begin{abstract}
The relation between symmetry breaking in non-commutative cut-off  field theories and transitions 
to inhomogeneous phases in condensed matter is discussed.
The non-commutative dynamics can be regarded  as an effective description of the mechanisms which 
lead to inhomogeneous phase transitions and their relation to the roton-like excitation spectrum.
The typical infrared-ultraviolet mixing in non-commutative theories contains the peculiar ingredients 
to describe the interplay between short and long distance particle interactions which is responsible 
for the non-uniform background and the roton spectrum both in bosonic and fermionic condensates.
\end{abstract}


\maketitle
The relation between symmetry breaking in quantum field theory and phase transitions in condensed matter is well known.
Spontaneous symmetry breaking in $\lambda \phi^4$ with a constant vacuum expectation of the field
corresponds to a Bose -Einstein condensation (BEC) and  there is a strong analogy between the cut-off field theory
and the hard sphere Bose gas description of superfluidity \cite{ccc}. Chiral symmetry breaking is the field theoretical version
of superconductivity and the ground state is described by a condensation of fermion-antifermion pairs corresponding to 
the Cooper pairs in the Bardeen-Cooper-Schrieffer (BCS) state \cite{miransky}.
The previous correspondences are usually restricted to the cases of constant
 order parameters, that is of transitions to uniform phases, because this
 guarantees the translational and rotational invariance of the field theory.
 
On the other hand, in condensed matter, due to particle interactions or to external experimental setups (for example a trapped
condensate), one considers transitions from homogeneous to inhomogeneous phases with non-constant order parameters.
For a bosonic system, according to \cite{braz}, the transition from homogeneous to inhomogeneous phases, with an order parameter
which for large distances is an oscillating function, is associated to a roton-like behavior of the excitation spectrum.
For the fermionic systems it is possible to build superconducting states, like the 
Larkin-Ovchinnikov-Fulde-Ferrell (LOFF) state \cite{loff},
with energy lower than the BCS state, where the Cooper pairs have a non-zero total momentum and then the fermionic condensate
is not uniform.

In \cite{hall} the application of non-commutative dynamics to the lowest Landau level, to quantum Hall effect
and to fluid dynamics has been discussed.
In \cite{gubser,noi1,noi2} it has been proposed that one can continue to follow the previous  analogies with
condensed matter also for transition to inhomogeneous phases if the quantum field theory is generalized to non-commutative coordinates.
In particular, the non-commutative generalization of $\lambda \phi^4$ shows \cite{gubser,noi1,bieten} that the spontaneous symmetry 
breaking occurs for  a non-uniform stripe phase and in the non-commutative Gross-Neveu (GN)  model there is an
inhomogeneous chiral symmetry breaking corresponding to  spin density waves \cite{noi2}.

In this letter we consider the symmetry breaking in cut-off non-commutative field theories as an effective approach to 
understand the general dynamical mechanisms which lead to inhomogeneous phase transitions and the correlated roton-like
excitation spectrum.
Indeed the non-commutative dynamics indicates that also for condensed matter systems: 

{\bf a)} there is a strong link between a roton-like excitation spectrum and a transition to an inhomogeneous background, 
which  confirms \cite{braz};

{\bf b)} for bosonic systems the non-uniform behavior and the correlated roton spectrum is due to non-local particle interactions
(also  induced by external setups), i.e., to the interplay between short and long distance effects;

{\bf c)} for fermionic systems there are: a  transition to inhomogeneous states with 
a pairing with total momentum $P \neq 0$; a roton dip in the excitation spectrum;
effects due to momenta non-equal to the Fermi momentum 
( this is mainly due to the infrared/ultraviolet (IR/UV) connection (for a review see \cite{iruv}));

{\bf d)} the observed phase transitions turn out to be  first order. 

\noindent
We shall first discuss the relation between the non-commutative 
scalar field theory and the roton excitation in BEC and  then we consider  fermionic systems.

On general grounds, in a bosonic condensate the roton spectrum  is due to a non-local interatomic
potential $V(\vec r - \vec r\,')$, with a  momentum dependent Fourier transform \cite{fey}. Indeed, the roton spectrum has been recently
obtained by considering a BEC with a ground state wave function
which takes explicitly into account the van der Waals interaction \cite{bec} or by
a dipole-dipole induced atomic interaction in trapped condensate \cite{dip1,dip2,dip3,dip4} 
or by considering a BEC close to the solid phase \cite{santi}.
Since the BEC with the local (pseudo)potential $\delta(\vec r- \vec r\,')$ is analogous to the spontaneous symmetry breaking
in  $\lambda \phi^4$ theory, one can assume that some relevant  physical effects due to the
non-local  repulsive  interaction can be described
by generalizing the self-interacting field theory  in such a way to introduce an effective non-local coupling.

\par A simple approach is to consider the non-commutative $\lambda \phi^4$ theory with action
\be
\label{uno}
S(\phi)=\int d^4 x \left (  {1 \over 2}\partial_{\mu} \phi ~ \partial^{\mu} \phi
- {1 \over 2} m^2 \phi ^2 -{  \lambda
\over 4!} \phi ^{4*}\right )
\ee
 where the star (Moyal)  product is defined by ($i,j =1,.,4$)
\bea
\phi ^{4*}(x)= \phi(x) * \phi(x) * \phi(x) * \phi(x) = \hspace{70 pt}&&\nonumber\\
\label{due}
\exp\left \lbrack{ i \over 2} \sum_{i<j}\theta_{\mu \nu} \partial^{\mu}_{x_i} \partial^{\nu}_{x_j}\right \rbrack
\Bigl ( 
 \left.
\phi (x_1) \phi (x_2) \phi (x_3) \phi (x_4)
\Bigr ) 
\right |_{x_{i}=x}&&
\eea
The ``deformation'' of the self-interaction term by the Moyal product gives a  
momentum dependent repulsive effect which is responsible,
as we shall see below,  for the roton spectrum and for the phase transition to an inhomogeneous background.

In \cite{noi1}  the spontaneous symmetry breaking 
for the theory in Eqs.  (\ref{uno}) and (\ref{due}) has been analyzed with the following results:

{\bf 1)} the transition occurs to a stripe phase where the order parameter is $\phi(\vec x) =A \cos \vec Q \cdot \vec x$;

{\bf 2)} $A,Q$ and the energy excitation $\omega(p)$ are fixed by minimizing the energy;

{\bf 3)} $\vec Q$ is orthogonal to $ \vec \theta$ and $Q$ is small for large $\theta$;

{\bf 4)} the excitation spectrum can be approximated by
\be
\label{tre}
\omega(\vec p)= p^2 + M^2(\vec p)
\ee
where the function $M(\vec p)$ will be discussed later.

The previous results, obtained by variational methods in the static limit,
represent the solutions of a set of self-consistent minimization equations of the effective potential in the Hartree-Fock approximation
for the parameters $A,Q,M$.
As discussed in detail in \cite{noi1}, since $Q$ is small, the inhomogeneous background is a smooth function of $x$ and then,
the  breaking of translational (smooth) and rotational invariance is approximated by a translational invariant propagator with a
momentum dependent mass term.

In the particular case $\theta_{ij}= \epsilon_{ijk}\theta^k$  with $\vec \theta = (0,0,\theta)$, 
$\vec Q = (Q/\sqrt{2}, Q/\sqrt{2},0)$  and large values of $\theta\Lambda^2$ ($\Lambda$ is the UV cut-off),
it turns out that
\be
\label{quattro}
\frac{Q^2}{\Lambda^2}=\left ( {\lambda \over {24 \pi^2 }}\right )^{1/2} \frac{1}{\theta\Lambda^2}
\ee
and that $M(\vec p)$, for small $p$, is given by
\be
\label{cinque}
M^2(\vec p)\left. \right|_{p\to 0} \sim \alpha 
+{ {\lambda} \over {6\pi^2 }}\frac{1}{|\vec p  \times \vec \theta|^2}
\ee
where $\alpha$ is a constant and  $\times$ indicates the usual vector product.

The peculiar behavior for small $p$ of the last term in the previous equation is due to the IR/UV connection
 of the non-commutative field theory and  gives a divergent mass term
in the IR region and a minimum of the irreducible two-point function. 
However the effective theory has a natural self-generated IR cut-off $Q$ where it is more correct to cut
the small momenta also because one is neglecting the phonon branch. Then the excitation spectrum, which is related to the
phase transition, should be correctly identified by Eqs. (\ref{tre}) and  (\ref{cinque}) with $p \geq Q$ and has a roton-like dip 
at a typical scale of order $Q$.

It should be clear that the Moyal-deformed term in Eq. (\ref{uno})  can mimic 
effective interactions which are non-local and globally repulsive. 
Then,  within this simple model, 
the previous results of the non-commutative theory can describe some 
interesting physical effects of the Bose -Einstein condensates
when the repulsive interaction is dominant and one focuses only 
on the rotonic part of the spectrum
because the spontaneous symmetry breaking with a single 
scalar field gives a gap.

For example, for a system trapped along the $z$ direction,
with a typical dimension L, where atoms interact by a dipole-dipole interaction, it has been shown 
by standard condensed matter methods \cite{dip1}, that
 the excitation spectrum has different features according to the value of the 
momentum  $\vec q$ in the $x-y$ plane: if $qL << 1$ there is a phononic behavior; 
if $qL >>1$, due to the dependence of the coupling strength on the momentum, there is a roton excitation.
All the dipoles are oriented along the $z$ axis and for a tight confinement one has (almost) a dipole system in the $x-y$ plane
with a repulsive dipole-dipole interaction.

We shall see that an analogous result on the roton spectrum can be obtained by the non-commutative $\lambda \phi^4$
theory.  Moreover, for a dipole-dipole interaction, there is another reason why the  system can be associated with a
non-commutative dynamics. According to \cite{bigatti} it easy to show that in the $x-y$ plane a
system of interacting dipole-like objects, compound by two bound opposite charge particles in a 
strong magnetic field $B$ along the $z$ axis, follows a non-commutative dynamics
\be
\label{sei}
[x,y] \simeq -i /B
\ee
Although this formal analogy with the usual dipole-dipole interaction should be considered in a restricted sense, mainly  because
 in the non-commutative dynamics the "dipole" dimension depends on its center mass momentum,
the previous discussion suggests that some features of the previous BEC with trapped atoms with magnetic moment
can be reproduced by considering
the $\lambda \phi^4$  field theory with  the non-commutative parameter
in the same direction of the dipole orientation
\be
\label{sette}
[x,y]= i \theta
\ee

Indeed, by  Eqs. (\ref{tre}), (\ref{cinque}) and (\ref{sette}), in the small $p$ region, 
it is easy to verify that the non-commutative $\lambda \phi^4$  theory, in the $x-y$ plane, 
has a roton-like spectrum, similar to Figure 1 of \cite{dip1} with a minimum at momentum $p_x^2+p_y^2=2Q^2$
(with $(p_x^2+p_y^2)L^2 >> 1$).  The roton spectrum of the non-commutative theory is associated to a spatially 
modulated background in the $x-y$ plane according to point {\bf 1)}, quoted above, and one expects
that a similar oscillating behavior should be present in the dipolar atomic BEC.
This aspect has not been analyzed in \cite{dip1}, but
an oscillating condensate was obtained by solving the Gross-Pitaevskii equation with laser induced 
dipole-dipole interaction in an elongated  cigar-shaped BEC \cite{dip3} and, in  \cite{dip2}, it has been shown
that, in the same geometrical configuration,  the dipolar interaction gives a roton dip in the excitation spectrum.

This dynamical correlation between roton-like spectrum and non-uniform background 
is not limited to dipole-dipole (induced) interactions. Indeed 
in \cite{santi} it has been suggested that a roton spectrum also occurs in 
a BEC close to a Mott-insulating phase with a macroscopic population of the states with 
momenta equal to the reciprocal lattice vector.
The above  examples confirm, in our opinion,  the suggestion given in \cite{braz} that the roton spectrum 
is correlated to an  inhomogeneous phase  
and the crucial dynamical ingredient is a non-local, globally repulsive interaction.

After the indications obtained for bosonic systems, let us now consider
the informations coming from non-commutative effective field theories for 
fermionic system with a transition from a homogeneous phase, where $<\bar \psi \psi>$ is
constant, to an inhomogeneous one, where $<\bar \psi \psi>$ is $\vec x$ dependent.
The simplest field theoretical approach to chiral symmetry breaking, corresponding to  superconductivity,
is the GN model with four fermion interaction described by the lagrangian
\be
\label{otto}
L(x)= i \bar \psi \partial \hskip -0.2 cm \slash \psi  + g (\bar \psi \psi)^2\; .
\ee

In 4 dimensions, for $g$ larger than some critical value, $g_c$,  the chiral symmetry is broken
and the dynamical generated mass is constant $m_0 \simeq g <\bar \psi \psi>_0$.
\par In \cite{noi2} the transition from homogeneous to inhomogeneous phase
has been obtained by generalizing the GN model to the non-commutative case with lagrangian
\be
\label{nove}
L(x)= i \bar \psi_\alpha \partial \hskip -0.2 cm
\slash \psi_\alpha  + g \bar \psi_\alpha * \psi_\alpha *\bar \psi_\beta* \psi_\beta
- g \bar \psi_\alpha * \bar \psi_\beta * \psi_\alpha* \psi_\beta\; .
\ee
For $g$ larger than some critical value, one finds again chiral symmetry breaking but, this time, in an inhomogeneous  phase
where the pair correlation function has a dependence on a total momentum, $\vec P$ of the ("Cooper") pair,
with $P \simeq  (1/\theta \Lambda ^2)$.
\par The order parameter turns out to be an oscillating function of $\vec x$ and one has the breaking of translational, rotational and chiral
invariance:
\be
\label{dieci}
<\bar \psi(x)  \psi(x) > =\left  ( 1+c\, P^2 {\rm cos} \left ( Px\right )\right ) <\bar \psi  \psi >_0\; ,
\ee
where$ <\bar \psi  \psi >_0$ is the constant order parameter of the commutative case and $c$ is a numerical constant.

Since the translational invariant breaking effects are small for large $\theta$,
in the analysis of the non-commutative GN model in  \cite{noi2} the static two-point function has been computed 
to order $O(P^4)$ and the  background,  $<\bar \psi(x) \psi(x)>$, turns out to be  a slowly changing function.
Then one can identify the quasi-particle spectrum as $\omega^2 (\vec k)=k^2+m^2(\vec k)$.
As in  the scalar case,
for small momenta, $k \geq P$, the quasiparticle spectrum
has  a leading contribution due to the IR/UV connection of the non-commutative theory, and the self-consistent
IR  and UV  behavior of the gap equation has been interpolated by
\be
\label{tredici}
m(\vec k) = m_\theta  \left [ 1+ \frac{g}{ \pi^2}\frac{1}{ |\vec k\times \vec\theta |^2}\right ]
\ee
where the constant $m_\theta$ has been fixed by the minimization of the energy and reduces to $m_0$ in the planar limit 
$\theta\Lambda^2\to\infty$, where the non-commutative effects disappear.
Also in this case the spectrum has a  roton-like dip in the plane orthogonal to $\vec \theta$ and one recovers 
the dynamical relation with  the non-uniform ground state \cite{braz}.

The previous field theoretical model is then analogous to a condensed  matter system with a non-local 
four-fermion interaction, with an inhomogeneous phase where the particle-hole (p-h) pairs  have a non-zero total 
momentum.  From this point of view the model mimics a non-s wave superconductor in a strong coupling regime.

A kind of  p-h instability has been originally proposed in \cite{over} for specific electronic materials 
(for a review see \cite{gruner}) and is known as spin density wave.
An analogous phase, known as chiral density wave \cite{cdw},  
has also  been analyzed in QCD at finite  density (for a review see \cite{raja,beppe}).
In \cite{tatsumi} it has been suggested the possibility of a color magnetic superconducting phase, 
above a critical density, by breaking the chiral symmetry in a chiral density wave background.

The results of the non-commutative models, where the anisotropic behavior is due to $\vec \theta$,
indicate that in condensed matter and in QCD systems {\it in analogous conditions}, i.e.,
with a pairing for  $\vec P \neq 0$ and in strong coupling regime, there is a transition to an
 inhomogeneous phase  and the excitation spectrum should have a small roton-like dip.

In the standard treatment of inhomogeneous superconductivity the anisotropic behavior of the gap
is usually associated only with a dependence on the angular variables of the quasiparticle momentum,
because its module is fixed at the Fermi momentum. 
In our case there is a  dependence on the momentum 
and a deformed Fermi surface \cite{nota}. 

Then, also for  fermionic condensates, 
the effective field theory approach suggests that a roton dip in the spectrum 
should be related to a non-uniform background. In fact this result 
has been for example obtained by analyzing the relation between chiral 
symmetry breaking and magnetism in \cite{tatsumi} with 
isoscalar and isovector chiral density wave background.
The lower energy spectrum is given by
\be
E_p^2=\vec p^{\,2} + M^2 + \frac{ \vec q^{\,2}}{4} - \sqrt{(\vec p \cdot \vec q)^2 + M^2 \vec q^{\,2}}
\ee
where $\vec q$ is the momentum of the chiral density waves. For $\vec p$ parallel to $\vec q$, $E_p$ has a minimum for
$p= \sqrt{q^2/4 - M^2}$. The condition $q^2/4 - M^2 \geq 0$ is fulfilled for all values of $q$ and $M$ which minimize the energy
in the range $\mu_1 \leq \mu \leq \mu_2$ of the chemical potential where the transition occurs.

Finally, the field theoretical  results indicate that the phase transitions to   inhomogeneous
condensates are first order and one expects similar behavior for the corresponding  condensed matter systems.
This conclusion could be a by-product of the self-consistent approach \cite{gubser,noi1,noi2}  but, at least for 
the scalar case, it has been confirmed by lattice calculations \cite{bieten}.
Points {\bf (a)-(d)} summarize the results of this letter, pointing out 
the general relation between a minimum in the self-energy and the 
inhomogeneous background, due to non-contact interactions, as the 
common underlying ingredient  shared by the various 
physical systems   and set-ups analyzed above.

\vspace{20 pt}
\begin{acknowledgments}
We acknowledge R. Jackiw, R. Casalbuoni, 
M. Baldo for useful suggestions and we are also grateful to 
G. Angilella, F. Cataliotti, M. Consoli for 
frequent discussions during the preparation of the paper and to G. Nardulli 
for reading the manuscript.
\end{acknowledgments}


\begin{thebibliography}{18}
\bibitem{nota} Indeed, the general scaling argument in \cite{pol} does not apply 
to a non-commutative theory with Moyal product.
\end{thebibliography}
\end{document}